\newcommand{\be}{\begin{equation}}
\newcommand{\ee}{\end{equation}}
\newcommand{\bea}{\begin{eqnarray}}
\newcommand{\eea}{\end{eqnarray}}
\newcommand{\bc}{\begin{center}}
\newcommand{\ec}{\end{center}}
\documentstyle[preprint,aps]{revtex}
\draft
\begin{document}
\title{Non-perturbative renormalization group approach
for a scalar theory in higher-derivative gravity}
\author{A. Bonanno$^{1}$ and D. Zappal\'a$^{2}$}
\address{
$^{1}$Louis Pasteur University, 3 Rue de l'Universit\'e,
67000 Strabourg, France
\\
and\\
$^{1}$Istituto di Astronomia, Universit\'a di Catania\\
Viale Andrea Doria 6, 95125 Catania, Italy\\
and\\
$^{2}$Dipartimento di Fisica, Universit\'a di Catania and 
I.N.F.N, Sezione di Catania\\ 
Corso Italia 57, 95129 Catania, Italy}
\maketitle
\vspace{1 cm}
\begin{abstract} 
\baselineskip 14pt       
A renormalization group study of a scalar theory coupled to gravity 
through a general functional dependence on the Ricci scalar in the 
action is discussed. A set of non-perturbative flow equations 
governing the evolution of the new interaction terms generated 
in both local potential and wavefunction renomalization is derived. 
It is shown for a specific model that these new terms play an important 
role in determining the scaling behavior of the system above the mass 
of the inflaton field. 
\\
\end{abstract}

\section{Introduction}
It is well know that modifications of Einstein's theory of gravity
are required in order to include quantum effects.
Although a consistent fundamental
theory of the spacetime near the Planck scale is still not available, 
one can consider General Relativity as a quantum effective field
theory \cite{adler,donoghue}. From this point of view one gives up the 
requirement of (perturbative) renormalizability to consider a more
general action that is a functional of any geometrical invariant $\Re$ 
which can be constructed from the principle of general covariance, 
$\Re=\{R, R_{\alpha\beta}R^{\alpha\beta}, C_{\alpha\beta\gamma\delta} 
C^{\alpha\beta\gamma\delta}, ...\}$, and of the matter field $\phi$ and 
their derivatives 
\be\label{azi}
S[\phi,g]=\int d^4 x \sqrt{-g}{\cal L}(\Re,\;\phi,\;\nabla^2\Re, 
\nabla^2\phi, \;\nabla^{2j}\Re,\;\nabla^{2j}\phi,...,)
\ee
This new theory is analogous to the Euler-Heisenberg lagrangian,
low-energy effective theory of the more fundamental 
microscopic QED.  
Due to the smallness of the ratio between the Newton constant and
the Fermi constant, this modification of Einstein's theory
is experimentally  indistinguishable from
the standard theory  at ordinary energy scales \cite{stelle}. 
However at higher energy scales 
\cite{ovrut} it has significant physical implications, mainly for the
inflationary scenarios \cite{staro}, nowadays  
tested by measuring the temperature fluctuations of the 
cosmic microwave background radiation (CMBR)
over large angular scales.  

Higher-derivative  gravity theories also arise 
through the coupling between a quantized field and the classical 
background geometry \cite{birrel}. In this case one adds new terms 
to the standard Einstein-Hilbert lagrangian. Those terms 
are general functions of $\Re$ and are needed to cancel the  
ultraviolet (UV) divergences at any given order in perturbation theory. 
This approach is useful to study the short-distance behavior of the
model, but it loses its validity if we are interested in the low energy 
behavior. 

In fact, the standard perturbative
approach based on the scaling property of the Green's
functions under a rescaling of the metric 
\cite{nelsonp} handles only a finite number of operators, those
which are important for the UV fixed point. 
In this way one cannot follow the evolution of  the irrelevant, 
i.e. non-renormalizable operators. 
Although the standard classification of the interactions
is given accordingly to the behavior of the renormalization group 
(RG) transformation near a
fixed point, we use a more general definition, calling
a coupling irrelevant, marginal
or relevant if, in cut-off units,
either it gets smaller, or it does not run, or it grows 
under a RG transformation. 
The irrelevant operators, even if they are not present in the bare
lagrangian, mix their evolution with the renormalized trajectory 
of the relevant couplings. Although near the gaussian UV fixed 
point their running is suppressed, they can behave in a quite 
different manner in other scaling regions thus deviating 
the renormalized flow. 

It is therefore important to trace the evolution of
all the couplings constants generated by the renormalization
procedure in order to decide what can be neglected and what 
cannot. If for example a new scale other than the cut-off 
is present in the problem, 
the scaling laws change near that scale and the
standard UV relevant interactions might not be sufficient
to describe the physics at the crossover. 
This happens in the $O(N)$ model where, in the non-symmetric phase, 
the presence of singularity in the beta functions
for the longitudinal components indicates that new infrared
(IR) relevant 
interactions, due to the Goldstone modes, appear after the condensate 
has formed \cite{lipo2}.  

Actually we deal with a similar situation when we move up in energy, 
going from an effective theory defined at some low energy scale, 
towards the high energy region. In fact integrating out the
heavy quarks contribution in the Standard Model generates
non-renormalizable vertices that are suppressed at the weak scale
\cite{ac} but that can show up when the heavy mass
threshold is reached \cite{apz}. 

Although there is no way to localize the energy
stored in the gravitational field, a local definition of 
a ``mass'' scale can 
always be given, even if no matter field acts as a source in 
Einstein's equations, i.e. $T_{\mu\nu}=0$. A possibility 
is to consider $m_G = \sqrt{|\Psi_2|}$
where $\Psi_2$ is the coulombian component of the 
Weyl tensor. For instance, in spherical symmetry 
it coincides with the square of the Weyl tensor, and for a Schwarzschild 
spacetime $\Psi_2=m/r^3$. Since $m^2_G \sim |R_{\alpha\beta\gamma\delta}|$
one can think of $m_G$ as the local effective mass producing the curvature
in that point. In this respect the gravitational field acts as an external 
field that influences the quantum dynamics within scales $l_G\sim m^{-1}_G$.

On the other hand, during the slowroll phase of the inflationary era, 
the only relevant ``mass'' scale is provided by $H=\dot{a}/a$,
$a(t)\sim a_0e^{Ht}$ is the scale factor,  and the 
quantum evolution of the fluctuations is stopped after crossing the horizon.
The subsequent classical behavior is therefore dependent
on the hierarchy between these two scales, the mass of the inflaton field, 
$m^2_{\phi}$, and the scalar curvature $R$. 

An important issue is related to the running of the conformal
coupling constant $\xi$ which couples directly the inflaton with gravity 
through the $\xi R\phi^2$ term in the action. 
The value of this coupling is determined in principle 
by the spectral index of the CMBR temperature fluctuations 
$n_s=1.17\pm 0.31$ \cite{cobe}, depending on the model of inflation considered. 
Although for some models negative values of $\xi$ are preferred 
\cite{salopek}, in perturbation theory with 
the minimal subtraction scheme one finds $\xi=1/6$ to be an infrared 
fixed point for the renormalized coupling \cite{nelsonp,odintsov}. 

The renormalization group approach used in Statistical Mechanics
is the best tool to study problems where many scales are
coupled together \cite{wilson}. 
In this paper we shall use the differential form of the RG transformation
formulated by Wegner and Houghton \cite{wegner}. Starting
from a bare action $S_k$ at the cut-off $k$ one first calculates 
$S_{k-\Delta k}$ in 
\be\label{azi1}
e^{-S_{k-\Delta k}[\phi]}=\int {\cal D}[\psi]e^{-S_k[\phi+\psi]}
\ee
by using the loop expansion, where $\psi$ and $\phi$ respectively
have non-zero Fourier 
components only in the momentum shells $k-\Delta k<p\leq k$ and 
$p\leq k-\Delta k$. The differential RG transformation 
is obtained by taking the limit of an infinitesimal shell
$\Delta k/ k \rightarrow \delta k / k$. 
The higher loop contributions in Eq. (\ref{azi1}) are suppressed
as powers of $\Delta k/ k$ in the limit $\Delta k \rightarrow 0$ 
for finite $k$
and an exact, non-perturbative, one-loop RG equation is obtained   
\be\label{wh}
k{d S_k[\phi]\over dk}=-{1\over 2}\langle \ln {\delta^2 S[\phi]
\over\delta\phi^2} \rangle +
{\langle {\delta S[\phi]\over \delta \phi} 
\Bigl ({\delta ^2 S[\phi]\over \delta \phi^2}\Bigr)^{-1}
{\delta S[\phi]\over \delta \phi}\rangle}
\ee
where the brackets indicates sum over the Fourier components 
within the shell. This functional equation rules the evolution of all 
the interaction terms that are generated in the renormalization
procedure. Despite the solution is not know in the general case
one can derive approximate, non-perturbative evolution equations
in terms of the gradient expansion \cite{fraser}, by writing
\be\label{gra}
S[\phi]=\int d^D x \sum_n U_n(\phi, \partial^{2n}\phi)
\ee
where $U_n$ is an homogeneous function of the field and of
its derivatives. In this way Eq. (\ref{wh}) decouples in a
set of infinite non-linear partial differential equation
for the $U_n$. 
The explicit construction of the RG transformation can be achieved by 
the blocking transformation or ``coarse-graining'' 
of the fields by means of a smearing function 
that introduces a sharp cut-off in the high momentum modes \cite{lipo1}. 
A perturbative construction of the coarse-graining approach has been firstly
discussed in a gravitational context in \cite{hu1} and proposed in the
formalism of the ``average action'' in \cite{me,per1}.
In particular in \cite{me}, 
by considering the contribution of the blocked potential $U_0(\phi)$ 
in Eq. (\ref{gra}), it has been shown that $\xi=1/6$ is not necessarely  an 
infrared fixed point for the massive theory coupled with gravity. 

An equivalent construction of the exact RG equations in the Wilsonian approach 
is presented in \cite{wett}. In particular in \cite{reu} 
the infrared behavior of the Einstein-Hilbert action is studied in the case of 
pure gravity.
 
In this paper we explicitly construct the RG transformation for the blocked 
potential $U_k$ and the wavefunction renormalization $Z_k$ which are general 
functions of both  $\phi$ and the curvature scalar $R$.
In particular we construct the RG traformation by means of an
$O(D)$ symmetric  smearing function ($D$ is the dimension of the spacetime). 
This will be achieved by introducing locally a momentum space \cite{parker}
and by working up to first non-trivial order in the expansion of the metric. 
The set of equations we derive generates the running of any interaction 
term of the form $g_{ij}R^i \phi^j$ in both $U_k$ and $Z_k$.

We study the RG flow in the two different scaling regions of the
ultraviolet and infrared domain.  
In particular the crossover in corrispondence of the mass gap is analyzed 
in a specific model by means of a numerical investigation of the
equations.
 
The organization of the paper is the following. In Sec.II 
the flow equations for $Z_k$ and $U_k$ are derived. 
In Sec. III we analitically study the behavior in the UV and 
in the IR region. 
In sec. IV we apply the method to a model by numerically 
integrating the flow equations. The results of the numerical analysis
are then displayed. Sec. V is devoted to the conclusions.

\section{renormalization group transformation}

In the following we suppose that the spacetime has a
well defined Euclidean section $\Omega$. 
The general form of the Euclidean bare action, quadratic in the 
derivatives of the scalar field $\phi$ and of the Ricci
scalar $R$ is  
\bea\label{1.1}
&&S_{\Lambda}[\phi,g]=\int d^{D}x\sqrt{g}
\{\frac{1}{2}Z(\phi,R)
\partial^\mu\phi \partial_\mu\phi+V(\phi,R)+\nonumber\\[2mm]
&&{1\over 2}W(\phi,R)\partial_\mu R\partial^\mu R
+Y(\phi,R) \partial_{\mu}\phi \partial^{\mu}R\}
\eea 
where $\Lambda$ is the ultraviolet cut-off of the theory,
$V$ is the local potential, while ${Z}, {Y}$ are general 
functions of $\phi$ and $R$. 
The action (\ref{1.1}) is equivalent to the following one 
\bea\label{1tris}
 &&S_{\Lambda}[\phi,g]=\int d^{D}x\sqrt{g}
\{-\frac{1}{2}{\cal Z}(\phi,R)
\phi \nabla^2 \phi+V(\phi,R)-\nonumber\\[2mm]
&&{1\over 2}{\cal W}(\phi,R) R \nabla^2 R
+{\cal Y}(\phi,R) \partial_\mu\phi \partial^{\mu}R
\}
\eea
provided
\be\label{1bis2}
Z=({\cal Z}\phi)',\quad W={\partial ({\cal W}R)\over \partial R},\quad
Y={\cal Y}+{1\over 2}R{\cal W}~'+
{1\over 2}\phi{{\partial{\cal Z}}\over{\partial R}}
\ee 
where $\nabla^2=\nabla^\mu\nabla_\mu$ is the Laplace-Beltrami 
operator, and the prime indicates derivation with respects to $\phi$.

In order to construct 
the renormalization group transformation 
in the Wilson-Kadanoff approach one first defines the 
average blocked field by coarse-graining the original field 
\be\label{1.2}
\phi_k (x)=\int d^{D}x'\sqrt{g} \rho_k(x,x') \phi(x')
\ee
where $\rho_k(x,x')$ is a smearing function which is constant 
in a given volume where the field is averaged, and it
is zero outside.
The blocked action is by definition
\be\label{1.3}
e^{-\widetilde{S}_{k}[\Phi,g]}=\int
D[\phi]\prod_{x}\delta(\phi_{k}(x)-\Phi(x))e^{-S_\Lambda[\phi,g]}.
\ee
In this way one ``averages out''  
the high energy degrees of fredoom
that are relevant for the short-distance behavior of the 
theory. 

Let us consider the decomposition of a generic function 
$f(x)$ of ${\cal L}^2(\Omega)$, in eigenfunctions of the Laplace-Beltrami
operator 
${\psi_{i}(x)}$,
$f(x)=\sum_i f_i \psi_i(x)$,
where the ${\psi_{i}(x)}$
represent an orthonormal base in this space 
($i$ is an integer if the manifold is compact). 
Once the spectrum of the operator $\nabla^2$ is known,
the sharp momentum smearing function defined in \cite{me} 
can then be used to explicitly construct the 
projector $\rho_k(x,x')$ in Eq. (\ref{1.2}). 
In particular, for a general spacetime 
one can introduce locally a
momentum space and use the $O(D)$ sharp-cutoff
smearing function as in \cite{lipo1}. This is 
a good approximation as long as the cut-off of the theory
is much above the scale of energy where the curvature becomes dynamically
relevant and, at the same time, it gives a precise meaning to the UV cutoff
$\Lambda$ introduced above.
The field is then split
\be\label{1.4}
\phi(x)=\vartheta(x)+\xi(x)=\sum_M \vartheta_m \psi_m(x)+\sum_N 
\xi_n \psi_n(x)
\ee
where $M$ and $N$ respectively indicate the slow and the fast variables;
more precisely $M$ is the set of index values $m$ labelling components 
with low momentum:
$p<k$ and $N$ is the set of index values $n$ and $n'$ 
corresponding to high momentum $k<p<\Lambda$, 
$k$ being a fixed momentum scale below the cutoff $\Lambda$.
The functional integration in Eq. (\ref{1.3}) yields
\bea\label{1.5}
&&{\widetilde S}_k[\Phi]=-\ln\int {\cal D}[\vartheta]{\cal D}[\xi]\prod_{x}
\delta (\phi_k(x)-\Phi(x))e^{-S[\vartheta+\xi]}=\nonumber\\
&&-\ln\int {\cal D}[\vartheta]\prod_M\delta(\vartheta_m-\Phi_m)
\int {\cal D}[\xi]{\exp}\Bigl\lbrace-S[\vartheta]
-{1\over 2}\sum_N F_n \xi_n-{1\over 2}
\sum_N\xi_n K_{n,n'} \xi_{n'} ...\Bigr\rbrace=\nonumber\\
&&-\ln\int {\cal D}[\vartheta]\prod_M\delta(\vartheta_m-\Phi_m)
{\exp}\Bigl\lbrace-S[\vartheta]+{1\over2}\sum_N F_n K_{n,n'}^{-1}F_{n'}
-\frac{1}{2}{\sum_N}\left (\ln K \right )_{n,n}\Bigr\rbrace=\nonumber\\
&&S[\Phi]-\sum_N F_n K_{n,n'}^{-1} F_{n'}\Big|_{\Phi}
+{1\over 2}{\rm Tr}~\ln K[\Phi]   
\eea
where
\be\label{1.6}
F={\delta S\over \delta\phi}\Big |_{\phi=\Phi},
\quad\quad K={\delta^2 S\over \delta\phi\delta \phi}\Big \vert_{\phi=\Phi}
\ee
In coordinate representation we get 
\bea\label{1.7}
&&F(x)={\delta S\over \delta \phi(x)}\Big\vert_{\Phi}
=-{1\over 2}{Z}'\;\partial_\mu \Phi \partial^\mu\Phi 
-{Z}\;\nabla^2\Phi
+ V'
\eea
and 
\bea\label{1.8}
&&K(x,x')={\delta^2 S\over 
\delta \phi(x)\delta \phi(x')}\Bigg\vert_{\Phi}=\nonumber\\[2mm]
&&\lbrace-{1\over 2}Z''\;\partial_\mu\Phi\partial^\mu \Phi
-Z'\;(\partial_\mu\Phi\partial^\mu + \nabla^2 \Phi)-Z\;\nabla^2+V''
\rbrace \delta(x,x')=\nonumber\\
&&\lbrace -{1\over 2}{Z}'\;\nabla^2 \Phi
-{Z}\;\nabla^2 + V''\rbrace\delta(x,x')
\eea
where 
\be\label{delta}
\delta(x,x')=
g^{-{1/4}}(x)\delta(x-x')g^{-{1/4}}(x')=\delta(x-x')/\sqrt{g(x)}
\ee
In deriving Eqs. (\ref{1.7},\ref{1.8}), we have supposed that, for our purposes 
the fluctuations of $R$ are negligible or, in other words, we have suppressed 
in the action the terms containing $\partial_\mu R$; as a consequence $W$ and 
$Y$, although in principle could be not vanishing, do not appear in 
Eqs. (\ref{1.7},\ref{1.8}) and in the following calculations.

The last equality in  
Eq. (\ref{1.8}) 
follows after some
manipulations, but the easiest way to verify it, is to compute the second
functional derivative from the action (\ref{1tris}), and  
use the relation for $Z$ in Eq. (\ref{1bis2}). 

In order to explicitly carry out the blocking procedure,
the functional form of the blocked action in Eq. (\ref{1.5}) 
is needed, and the usual assumption is to mantain the same form of the 
starting action; thus, according to the restriction to constant $R$, the 
blocked action reads 
\bea\label{1.9}
&&\widetilde S_{k}[\Phi]=
\int d^D x\sqrt{g}\Biggr({1\over 2} \widehat {Z}_k(\Phi,R)~ \partial_\mu\Phi
\partial^\mu\Phi + U_k(\Phi,R)
+O(\nabla^4)\Biggr)
\eea
The function
$\widehat {Z}$ is generated by performing  
the path integral in Eq. (\ref{1.5}) for 
non constant field configurations 
\be\label{1.10}
\Phi(x)=\Phi_0+\widetilde\phi(x)
\ee
By expanding the action in Eq. (\ref{1.9}) 
up to second order in $\widetilde\phi$, one can single out the contribution 
to the blocked potential $U$ and to  the wavefunction renormalization 
$\widehat Z$, as it has been shown 
in \cite{fraser} and in \cite{lipo2}. 
In a general spacetime this procedure might be problematic
because the effective potential is in general ill defined
when the spacetime dynamics becomes relevant.  
For this  purpose it is convenient to consider energy scales
much higher than the characteristic curvature scale 
$k^2\ll R$, and as anticipated above, 
to introduce locally a momentum space \cite{parker}
and to use an $O(D)$ symmetric smearing function in Eq. (\ref{1.2})
to evaluate Eq. (\ref{1.5}). In fact, although several techniques
are available in the literature \cite{barvi} to compute the functional 
determinant in Eq. (\ref{1.5}), they are often combined with
dimensional regularization and  minimal subtraction prescription.
Instead, the explicit use of a sharp cut-off regulator
allows a clear separation between the UV ad the IR domain of the theory.

Let us briefly outline the computation of the blocked action. 
It is convenient at this point to introduce a covariant notation:
for any operator $A$ the Lorentz invariant operator $\overline A$ is
defined by ${\overline A}(x,x')\equiv g^{1/4}(x) A(x,x')g^{1/4}(x')$. 
In particular for the trace in Eq. (\ref{1.5}), properly written in 
coordinate representation, we get 
\be\label{tra}
{\rm Tr}\ln K=\int d^D x\sqrt{g(x)}\ln K(x,x) =
\int d^D x \ln \overline K (x,x)\equiv {\overline {\rm Tr}}~{\overline K}
\ee
where the definition of the logarithm as power series of operators 
has been used.
Let us examine the various pieces entering $\overline K$. 
According to the above definition 
$g^{1/4}(x) \nabla_x^2 \delta(x,x') g^{1/4}(x')=
\overline \nabla_x^2 \delta(x-x')$
where Eq. (\ref{delta}) has been used and
\bea\label{nabb}
&&{\overline\nabla}^2=g^{\mu\nu}(x)\partial_{\mu\nu}^2+b^{\mu}(x)\partial_\mu
+a(x)\nonumber\\[2mm]
&&b^{\mu}(x)=g^{-1/2}(\partial^\mu\sqrt{g})+2g^{1/4}
(\partial^\mu g^{-1/4})+(\partial_\nu g^{\nu\mu})\nonumber\\[2mm]
&&a(x)=g^{-1/4}(\partial_\mu\sqrt{g})
(\partial^{\mu}g^{-1/4})+
g^{1/4}\partial_\mu\partial^{\mu}g^{-1/4}.
\eea
Analogously
\be\label{stagarg}
g^{1/4}(x) Z'(\nabla_x^2 \Phi) \delta(x,x') g^{1/4}(x')=
g^{-1/4}(x) Z'~\overline \nabla_x^2 (g^{1/4} \Phi) \delta(x-x')
\ee
Thus, making the replacements
\be\label{repl}
\Phi(x) g^{1/4}(x) \rightarrow \Phi(x),~~~~~~~~~~~~
V(x) g^{1/2}(x) \rightarrow V(x)
\ee
we finally get 
\bea\label{erc}
&&\overline K(x,x') =
\lbrace-{1\over 2}{Z}'\;{\overline\nabla}^2(\Phi)
-{Z}{\overline\nabla}^2 +V''
\rbrace\delta(x-x')
\eea
where the prime now indicates the derivative with respect to the new scalar 
field of Eq. (\ref{repl}).
\par 
Note that we could have obtained Eq. (\ref{erc}), making the replacements 
of Eq. (\ref{repl}) in the original action and taking second functional 
derivatives with respect to the new scalar field.
Since the following developments are naturally expressed in terms of
the replaced  variables in Eq. (\ref{repl}), we shall keep on using the
new field and potential, but for simplicity we shall not change the notation.
\par
We can now introduce normal 
coordinates $\{y^\alpha\}$ around the  
point $x'$ corresponding to $y^\alpha=0$, 
expand the expressions in Eqs. (\ref{nabb}) and (\ref{erc}) up to 
some order and invert the propagator as shown
in \cite{parker} and \cite{hu2}
where they fixed $Z=1$ at the bare level.
However for our purposes 
it is more convenient to 
work out separately the expansions for the metric and the field $\Phi$ 
since in principle 
they involve different scales. 
We expand the metric tensor up to second order in its derivatives
therefore writing 
\bea\label{1.11}
&&g^{\mu\nu}(y)=\delta^{\mu\nu}+{1\over 3}{{{R^{\mu}}_{\alpha}}^\nu}_\beta 
y^\alpha y^\beta+O(\partial^3 g)\nonumber\\[2mm] 
&&g=1-{1\over 3}R_{\alpha\beta}y^{\alpha}y^{\beta}+O(\partial^3 g)
\eea
This truncation is consistent with the choice of neglecting $\partial_\mu R$,
that would appear starting from the third order in the metric expansion.

In order to compute $\widehat Z$ we first expand Eq. (\ref{erc}) 
around $\Phi(y)=\Phi_0+\widetilde\phi(y)$
up to second order in $\widetilde\phi$ obtaining
${\overline K}={\overline K}_0+{\overline K}_{\widetilde\phi}
+{\overline K}_{\widetilde\phi^2}$
where ($Z_0$ and $V_0$ are used for constant scalar field $\Phi_0$
and  $\Box=\delta^{\mu\nu}\partial_\mu\partial_\nu$ )
\bea\label{1.15}
&&{\overline K}_0(y,0)=
\lbrace -{Z}_0\Box+V''_0-{R\over 12}(2Z_0+\Phi_0 Z_0')
\rbrace\delta(y)\nonumber\\[2mm]
&&{\overline K}_{\widetilde\phi}(y,0)=
\lbrace -{1\over 2}{Z}_0'(\Box{\widetilde\phi}+2{\widetilde\phi}\Box)+V_0'''{
\widetilde\phi}
-{R\over 12} (3Z_0'+\Phi_0Z_0''){\widetilde\phi}\rbrace\delta(y)\nonumber\\[2mm]
&&{\overline K}_{\widetilde\phi^2}(y,0)
=\lbrace-{1\over 2}{Z}_0''({\widetilde\phi}\Box{\widetilde\phi}+
{\widetilde\phi}^2\Box)+{1\over 2}V_0''''{\widetilde\phi}^2
-{R\over 24} (4Z_0''+\Phi_0 Z_0'''){\widetilde\phi}^2\rbrace\delta(y)
\eea
In deriving these expressions we have taken into account that 
$\overline\nabla^2$, when applied to a scalar quantity, becomes 
$\overline\nabla^2
=(\Box+(1/3){{{R^{\mu}}_{\alpha}}^\nu}_\beta 
y^\alpha y^\beta \partial^2_{\mu\nu}-
(1/3){R^{\mu}}_\alpha y^\alpha\partial_\mu +{1\over 6}R)=
\Box+R/6$.
We rewrite Eq. (\ref{tra}) using   
\bea\label{1.16}
&&{\overline {\rm Tr}}\ln {\overline K}={\overline {\rm Tr}}\ln {\overline K}_0+
\sum_{n=1}^{\infty}
{(-1)^{n+1}\over n} {\overline {\rm Tr}} \Bigl \{ {\overline K}^{-n}_0
\left ( {\overline K}_{\widetilde\phi}+{\overline K}_{\widetilde\phi^2}
\right )^{n}  \Bigr \} 
\eea
One therefore also expands the expression (\ref{1.9}) up to second order 
in $\widetilde\phi$ and compare it with Eq. (\ref{1.5}), obtaining
\bea\label{nuova}
&&\widetilde S_k[\Phi]=
\widetilde S_k^0+
\widetilde S_k^1+
\widetilde S_k^2=S_\Lambda^0
+S_\Lambda^1+S_\Lambda^2+\nonumber\\[2mm]
&&{1\over 2}{\overline {\rm Tr}}\ln {\overline K_0}+
{1\over 2}{\overline{\rm Tr}}~{\overline K_0}^{-1} 
{\overline K}_{\widetilde\phi}+
{1\over 2}{\overline{\rm Tr}}~{\overline K_0}^{-1} 
{\overline K}_{\widetilde\phi^2}
-{1\over 4}{\overline{\rm Tr}}\big\lbrace{\overline K_0}^{-1}
{\overline K}_{\widetilde\phi}{\overline K}_0^{-1} 
{\overline K}_{\widetilde\phi}\big\rbrace
+O({\widetilde\phi}^3)
\eea
where $S^0, S^1, S^2$ are, respectively,  the zeroth, the first and the
second order terms in the functional expansion of the action in $\widetilde
\phi$. 
The zeroth order gives the blocked potential 
\be\label{pripo}
\widetilde S_k^0=S^0_\Lambda[\Phi_0]+{1\over 2}{\overline {\rm Tr}}\ln{
\overline K_0}
\ee
wherefrom 
(in the following we use the notation 
$\int_p=\int_{k<|p|<\Lambda}{d^D p\over (2\pi)^D}$)
\bea\label{pot}
&&U_k=V_0+{1\over 2}\int_p
\ln\big(Z_0p^2+V''_0-{R\over 12}(2Z_0+\Phi_0 Z_0')\big ).
\eea
In deriving the last relation we have evaluated the 
trace  in Eq. (\ref{pripo}) in the Riemann frame. Since the integrand 
is calculated in $\Phi_0$ and $\overline K_0$ is diagonal in the local momentum
representation, a volume term can be factored out, yielding a covariant 
equality between constant quantities. The blocked potential 
$U(\Phi)$ is obtained
by replacing the constant field $\Phi_0$ with $\Phi(x)$.
It should also be observed that Eq. (\ref{pot}) for $k=0$ gives the 
effective potential, obtained by taking the coincidence limit in the 
fluctuation determinant and by retaining only the first order terms in the 
Schwinger-De Witt proper time expansion \cite{sergio}. 

A similar procedure must be used for selecting the contribution of 
$\widehat {Z}_k$ 
in the other terms in the  expansion (\ref{nuova}).
In this case, for computing the traces, according to \cite{fraser}
we retain  
at the same time 
operators in the momentum and in the coordinates representation,
as long as the operators in one representation are on the right 
side of the operators in the other representation. In the terms  
where the required ordering is not fulfilled, we
employ commutation rules for $p$-dependent and $x$-dependent terms
in order to move the $p$-dependence on the left of the $x$-dependece
and disentangle the volume integration from the $p$ integration. 
The main commutation rule, from which all the others can be deduced 
with no ambiguity, is 
\be\label{1.17}
[\Phi(x),p_\mu]=-i \partial_\mu \Phi(x) 
\ee
Thus we get
\bea\label{equa}
&&{\overline {\rm Tr}}~{\overline K_0}^{-1}{\overline K}_{\widetilde\phi}=
\int d^D y \int_p {1\over K_p}
\big( Z_0'p^2+V_0'''-{R\over 12} (3Z_0' +\Phi_0 Z_0'')
\big )\widetilde\phi(y)\nonumber\\[2mm]
&&{\overline {\rm Tr}}~{\overline K_0}^{-1}{\overline K}_{\widetilde\phi^2}=
\int d^D y \int_p {1\over K_p}
\Biggl ( {1\over 2} Z_0''(\partial_\mu \widetilde\phi(y) \partial^\mu
\widetilde\phi(y)+p^2 \widetilde\phi^2(y))+
\nonumber\\
&&{1\over 2}V_0''''\widetilde\phi^2(y)
-{R\over 24} (4Z_0'' +\Phi_0 Z_0''')\widetilde\phi^2(y)
\Biggr )
\eea
where
\be\label{prpg}
K_p=Z_0~p^2+V_0''-(R/12)(2Z_0+\Phi_0 Z_0')
\ee
and, after having used the commutation relation (\ref{1.17}) for 
computing ${\overline{\rm Tr}}\ln({\overline K_0}^{-1}
{\overline K}_{\widetilde\phi}{\overline K}_0^{-1}
{\overline K}_{\widetilde\phi})$ 
we eventually obtain the following renormalization group equations
\bea\label{rgind}
&&U_k=V+{1\over 2}\int_p {\ln}(K_p)\nonumber\\
&&{\widehat Z}_k= {Z}
+{1\over 2}\int_p\Bigl( {Z'' \over K_p} -
{2 {Z}' K_p ' \over K_p^2} +
{{Z} (K_p ')^2 \over K_p^3}+ 
{ Z Z' p^2
~K_p ' \over K_p^3} -
{ Z^2 p^2
~(K_p ')^2 \over K_p^4} \Bigr)
\eea

Note that there is no contribution from the
term $FK^{-1}F$ in Eq. (\ref{1.5}) in the gradient expansion
as it can proved by direct calculation. The reason is that
in this term no volume integration occurs and one obtains  
the fourier transform of $\widetilde\phi$
which is constrained to have $p<k$. Since the momentum integration
is performed for $p>k$ this term identically vanishes \cite{lipo2}.

These equations have been obtained in the 
independent mode approximation where one computes the blocked
action by an independent integration of  the degrees of freedom between the
UV cut-off $\Lambda$ and $k$. In the limit $k\rightarrow 0$ one recovers
the effective action. In particular if we neglect the field dependence
in Z, therefore reducing it to a running coupling $\eta$, 
the previous equations read
\bea\label{rgind1}
&&U_k=V+{1\over 2}\int_p {\ln}
{\eta p^2+V''-(R/6)\eta)\over (\eta p^2+V'')|_{\Phi=R=0}}\nonumber\\
&&{\widehat Z}_k= \eta
+{1\over 2}\int_p\Bigl( 
{\eta (K_p ')^2 \over K_p^3} -
{ \eta^2 p^2
~(K_p ')^2 \over K_p^4} \Bigr).
\eea
where now $K_p=\eta p^2+V''-(R/6)\eta$ and the 
cosmological constant that
corresponds to the constant part of $V$ has been modified in
order to include the field independent part in the logarithm
in Eq. (\ref{rgind1}).

As an example let us consider the standard
$\lambda \phi^4$ bare scalar self-interacting theory in curved spacetime 
\bea\label{scaa}
&&{\cal L}=
{1\over 2}\eta \partial_\mu\Phi\partial^\mu\Phi+
{1\over 2}m^2\Phi^2+{1\over 2}\xi R\Phi^2
+{1\over 4!}\lambda\Phi^4
\eea
the RG equations (\ref{rgind1}) for 
$\widehat Z$ can be integrated up to $k=0$ and one obtains
in $D=4$ the finite contribution to the wavefunction renormalization 
function
\be\label{va}
{\widehat Z}_{k=0}=\eta+{1\over 192\pi^2}
\frac{\lambda^2\Phi^2}
{m^2+{\lambda\over 2} \Phi^2+(\xi-{1\over 6})R}
\ee
which coincides with the result in \cite{sergio}.
\par 
We want to remark that, due to the replacements in Eq. (\ref{repl}),
the couplings and the field appearing in the bare lagrangian do not coincide
with the ones appearing in the evolution equations, although, in case of
small curvature the difference becomes practically negligible. 
However it is formally correct  to expand in terms of 
running couplings the modified potential defined in Eq.(\ref{repl}) and not 
the bare lagrangian as in Eq. (\ref{scaa}).
\section{non-perturbative flow equations}
The previous equations have been obtained by eliminating 
the degrees of freedom between $\Lambda$ and $k$. 
The non-perturbative RG equations in Wegner-Houghton's 
formulation follow if one  infinitesimally lowers the running cut-off 
$k\rightarrow k-\Delta k$ thus retaining the contribution
of the modes which have been previously integrated 
in the infinitesimal momentum shell.
In practice the dependent mode improved
approximation is obtained by taking the derivatives with respect to $k$
of Eq. (\ref{rgind}) rewritten for a general potential and wavefunction 
renormalization. One obtains the following set of non-linear partial 
differential equations for the evolution of $\widehat Z$ and 
$U$ (for simplicity the subscript $k$ is omitted)
\bea\label{rgdep}
&&k{d U \over d k}= -a_D k^D{\ln}
{\widehat Z k^2+U''-(R/12)(2
\widehat Z+\Phi \widehat Z')\over 
(\widehat Z k^2+U'')|_{\Phi=R=0}}\nonumber\\[2mm]
&&k{d \widehat Z\over dk}=-{a_D k^D\over K}\Bigl( \widehat Z'' -
{2 {\widehat Z}' K ' \over K} +
{{\widehat Z} (K ')^2 \over K^2}+ 
{\widehat Z \widehat Z' k^2
~K ' \over K^2} -
{{\widehat Z}^2 k^2
~(K ')^2 \over K^3} \Bigr)
\eea
where $a_D=1/2^D\pi^{D/2}\Gamma(D/2)$ and
\be
K=\widehat Z k^2+U''-(R/12)(2
\widehat Z+\Phi \widehat Z')
\ee
These equations rule the evolution
of all coupling constants generated by the renormalization
procedure, and have been derived without any assumption on the
functional form of the potential and the wavefunction
renormalization. 
For instance, if $U$ and $\widehat Z$ are analytic functionals we write
\bea\label{1bis}
&&{\widehat Z}=
\sum_{i,j=0}^{\infty} {1\over i!j!} h_{ij}(k)\Phi^i R^j,\quad\quad
\quad\quad U=\sum_{i,j=0}^{\infty} {1\over i!j!} g_{ij}(k)\Phi^i R^j.
\eea
The beta functions for the 
couplings in Eq. (\ref{1bis}) are 
\bea\label{beta}
&&k{dg_{ij}\over dk} = {\partial^{i+j}U\over \partial^i\Phi
\partial^j R}|_{vac}\nonumber\\[2mm]
&&k{dh_{ij}\over dk}={\partial^{i+j}Z\over \partial^i\Phi
\partial^j R}|_{vac}
\eea
and one obtains an infinite dynamical system that can be conveniently 
truncated to some order to obtain an approximate solution of Eq. (\ref{rgdep}). 
Let us for instance neglect the field dependence in 
$\widehat Z$ using the improved version of Eqs. (\ref{rgind1})
in $D=4$ (the notation $\widehat Z=\eta_k$ is adopted analogously 
to Eqs.  (\ref{rgind1}) )
\bea\label{polipo}
&&k{d U \over d k}= -{k^4\over 16\pi^2}~~{\ln}
{\eta_k k^2+U''-(R/6)\eta_k \over 
(\eta_k k^2+U'')|_{\Phi=R=0}}\nonumber\\[2mm]
&&{d \ln \eta_k \over d \ln k} = -{k^4\over 16\pi^2}
{(U''')^2(U''-(R/6)\eta_k)\over 
(\eta_k k^2+U''-(R/6) \eta_k)^4}
\eea
It is interesting to employ these equations in order
to study the behavior of the couplings in the model (\ref{scaa}) 
which corresponds to a specific truncation in Eq. (\ref{1bis}).
 
We define the running parameters 
$m^2_k=g_{20}$, $\lambda_k=g_{40}$, $\xi_k=g_{21}$, $\eta_k=h_{00}$
and, as usual, the renormalized parameters 
$m_R, \lambda_R, \xi_R, \eta_R$ are obtained in the limit $k\to 0$
and, in the broken phase where $\langle \phi(x) \rangle=v$, 
$m^2_R=g_{20}(k=0)+\lambda_R v^2/2$.

By fixing the vacuum in Eq. (\ref{beta}) at the values 
$\langle \phi(x) \rangle=v$ and the Ricci scalar at the value $R={\cal R}$
we obtain the following $\beta$-functions 
\bea\label{siste}
&&k{d m_k^2\over dk}=-{k^4\over 16\pi^2} f_k\big(\lambda_k 
-f_k \lambda_k v \big)
\nonumber\\[2mm]
&&k{d \lambda_k\over dk}={k^4\over 16\pi^2}f_k^2\big(3\lambda_k^2
+6 f_k\lambda_k^4v^4+12\lambda_k^3 v^2 \big)
\nonumber\\[2mm]
&&k{d \xi_k\over dk}={k^4\over 16\pi^2}f_k^2\lambda_k
(\xi_k-\eta_k/6)\big(1-f_k2\lambda_k v^2)\nonumber\\[2mm]
&&k{d \tau_k\over dk}={k^4\over 16\pi^2}v \lambda_k^2 f_k^2\big(3
-2 f_k\lambda_k v^2 \big)
\nonumber\\[2mm]
&&k{d \sigma_k\over dk}={k^4\over 16\pi^2} v \lambda_k f_k^2
\big(\xi_k-{\eta_k \over 6} \big)
\nonumber\\[2mm]
&&\gamma_k={d \ln \eta_k \over d\ln k}=-{k^4\over 16\pi^2}f_k^4
\lambda_k^2 v^2\big (m^2_k+(\xi_k-\eta_k/6){\cal R}+\lambda_k v^2 /2\big)
\eea
where
\be\label{effek}
f_k^{-1}=\eta_k k^2+m_k^2+\lambda_k v^2/2+(\xi_k-\eta_k/6){\cal R}
\ee
From the above equations it is evident that the presence of a nonvanishing
expectation value of the scalar field $v$ has the consequence of generating
new couplings $\tau_k$ 
for the operator $\Phi^3$
and $\sigma_k$ for $\Phi R$.  In principle the location of 
the vacuum should be determined after 
solving the equation of the minimum for the effective potential, 
in the improved scheme. This amounts to solve the system (\ref{siste})
coupled to the equation of the minimum for the running potential.
Here things are more complicate since 
in higher-derivative gravitational theories 
the vacuum  does not necessarely coincide with the ${\cal R}=0$
configuration \cite{ovrut}. In the following we just consider $v$
and ${\cal R}$ as adjustable parameters.

We now discuss the scaling behavior of the couplings in  the two extreme
scaling limits: ultraviolet and infrared. In the UV domain we neglect 
all scales with respect to $k$, obtaining
\bea\label{sisteuv}
&&k{d m_k^2\over dk}=-{k^2\lambda_k\over \eta_k 16\pi^2} 
\nonumber\\[2mm]
&&k{d \lambda_k\over dk}={ 3\lambda_k^2\over \eta_k^2 16\pi^2} 
\nonumber\\[2mm]
&&k{d \xi_k\over dk}={\lambda_k (\xi_k-\eta_k/6)\over \eta_k^2 16\pi^2}
\nonumber\\[2mm]
&&k{d \tau_k\over dk}={3\lambda_k^2 v\over \eta_k^2 16\pi^2}
\nonumber\\[2mm]
&&k{d \sigma_k\over dk}={v \lambda_k \over \eta_k^2 16\pi^2} 
\big(\xi_k-{\eta_k \over 6} \big)
\eea 
while the anomalous dimension vanishes as 
\be\label{amol}
\gamma_k\sim O({1\over k^4})
\ee
as it should be. This result coincides with the findings in 
\cite{nelsonp,odintsov} for the $\beta$-functions in the UV domain. 
However once $k$ becomes much smaller than $m_k$, the scaling laws
change. Suppose first that the following hierarchy holds
\be\label{mash}
m^2_R \gg {\cal R}
\ee
then, below the mass gap the renormalized
flow is arrested and in $k=0$ the flow always reaches a 
(completely trivial) fixed point. 
In particular this shows that the findings in \cite{me} 
that $\xi_R=1/6$ is not necessarely an IR attractor, 
is still valid when the contribution of $\eta_k$ is considered. 
At this level of the approximation we note that $\xi_k=1/6$ 
is a fixed point only if $\eta_k=1$, statement which is not true in
the IR domain for the broken phase.  
In the symmetric phase, instead if we fix $\eta_k$ at the bare level 
to one it does not change and  $\xi_k=1/6$ is a fixed point for the
system (\ref{siste}). However, as we shall see in the
next section, when the contribution of the 
irrelevant operators is included, this fixed point disappears.
These conclusions should apply for any $D$
as one can see by direct calculation of the $\beta$-functions
derived from Eq. (\ref{polipo}). We also note that the gaussian fixed point is 
always present in any dimension.
  
When Eq.(\ref{mash}) is not satisfied  the structure of the vacuum 
can be different. In fact in general the curvature competes to determine the
sign of the gravitational running mass $\partial^2_\Phi U$ and it can lead
the sistem to a new phase. In fact, by looking at 
Eq. (\ref{effek}) we note that around
\be\label{insta}
\eta_k k^2 \sim (\eta_k/6-\xi_k){\cal R}-m^2_R
\ee
the equations in (\ref{siste}) become unstable because of the
presence of poles in the $\beta$-functions signaling that 
a new phase sets and that new IR relevant interactions are 
needed to describe the renormalized system. 

One might worry that around $k^2\sim R$ the local momentum
expansion is not reliable. This is not the case. In fact
for the Einstein Universe we have \cite{ab} the following 
exact finite-difference RG equation for the local potential
\be
U_{n-\Delta n, k-\Delta k}-
U_{n,k}=-{k \over 4\pi^3 a^3}n^2
\ln (a^2 k^2 +a^2\partial^2_\Phi U(\Phi, a^2)+n^2
-1)\Delta n \Delta k
\ee
where $n$ is the quantum number associated with the 
smearing on the spatial sections with topology $S^3$, $k$ is a cut-off
in the Euclidean ``time'' direction and $a^2=6/R$ is the radius of $S^3$. 
We find also in this case the appearance of a singular behavior 
at some scale $n$ if the ``restoring force'' 
$\partial^2_\phi U(\Phi, a^2)$ becomes negative. 

The physical reason of this instability is that the saddle-point expansion 
in Eq. (\ref{1.5}) has been performed by considering perturbations
around homogeneus configurations which do not  dominate the
path-integral in Eq. (\ref{1.5}) on scales smaller than the 
symmetry-breaking scale. 
In this case a more refined computation is needed, and the 
naive gradient expansion cannot be applied.

\section{the crossover}

The study of the crossover is instead more complicated and one has 
to integrate numerically the flow equations by keeping the contribution
of the ``irrelevant'' operators as well. 
In order to deal with a more tractable problem, we employ a 
truncation in the expressions (\ref{1bis}) of the local potential and the
wavefunction renormalization function, by keeping up to dimension-6 operators.
We thus consider the following action defined at the scale $k$
\bea\label{3.1}
&&S=\int d^D x\sqrt{g}\Bigl\{ \epsilon_0+\epsilon_1 R+{\epsilon_2\over 2}R^2
+{\epsilon_3 \over 3!}R^3
+{1\over 2}\eta_k\partial_\mu\Phi\partial^\mu\Phi+
{1\over 4}\alpha_k\Phi^2\partial_\mu\Phi\partial^\mu\Phi\nonumber\\[2mm]
&&+{1\over 2}\beta_k R \partial_\mu \Phi\partial^\mu\Phi+
{1\over 2}m_k^2\Phi^2+{1\over 2}\xi_k R\Phi^2
+{1\over 4!}\lambda_k\Phi^4+{1\over 2!2!}\chi_k R^2\Phi^2
+{\zeta_k \over 4!}R\Phi^4 +{1\over 6!}\omega_k\Phi^6 \Bigr\}
\eea
where $\epsilon_0=m^2_{PL}{\Lambda_{CS}}/8\pi$,
$\epsilon_1=-m^2_{PL}/16\pi$, 
$m_{PL}$ is the Planck mass, $\Lambda_{CS}$ is the cosmological constant.
Note that the remark at the end of sect. II also applies for the more general 
action (\ref{3.1}).

In order to avoid problems with triviality we keep the UV cut-off 
$\Lambda$ fixed. It is convenient to work in running cut-off units, 
by introducing the dimensionless variables 
\be\label{2.1}
x=\Phi k^{-(D-2)/2},\;\;\; y=R k^{2-D}
\ee
and calling $u,z$ respectively the dimensionless potential and wave 
function renormalization, we have the expansions
\bea\label{poze}
&&{z}=
\sum_{i,j} {1\over i!j!} z_{ij}(k)x^i y^j,\quad\quad
\quad\quad u=\sum_{i,j} {1\over i!j!} u_{ij}(k)x^i y^j
\eea
where one can easily find the proper correspondence among $z_{ij}(k)$,
$u_{ij}(k)$ and the various couplings appearing in Eq. (\ref{3.1}).
If we define
\be\label{2.2}
{\cal A}_t(t,x,y)=z(t,x,y)+\partial_x^2 u(t,x,y)-y(2z+x\partial_x z)/12
\ee
the renormalization group equations for $u$ and $z$ can be 
rewritten 
in terms of the new variables in the following way 
\bea\label{2.3}
&&{du\over dt}={(D-2)\over 2}(x\partial_x u+2y\partial_y u)-Du
-a_D{\ln}({{\cal A}_t \over {\cal A}_t|_{x=y=0}})\nonumber\\[2mm]
&&{dz\over dt}={(D-2)\over 2}(x\partial_x z+2y\partial_y z)-\nonumber\\[2mm]
&&a_D{\cal A}^{-1}_t \Bigl( \partial_x^2 z-2{\cal A}_t^{-1}\partial_x z
\partial_x {\cal A}_t+{\cal A}_t^{-2}z[(\partial_x {\cal A}_t)^2+
\partial_x z\partial_x {\cal A}_t]
-{\cal A}_t^{-3}z^2(\partial_x {\cal A}_t)^2
\Bigr).
\eea 
We have rescaled the momentum variable, by defining 
$t=\ln(k/\Lambda)$ so that we generate the renormalized flow 
by lowering the cut-off $\Lambda\rightarrow 
\Lambda e^t$, with $t\leq 0$. 

We insert Eq. (\ref{poze}) in Eq. (\ref{2.3}) and we evaluate 
the derivatives in the symmetric vacuum  $\Phi=0$ at 
non-zero curvature $R={\cal R}$ (the corresponding 
dimensionless value of $y$ is indicated with $r$:
$r={\cal R} k^{2-D}$ ), obtaining in D=4 
the system of ordinary differential equations displayed in the
appendix. For simplicity we shall not discuss here the symmetry broken 
phase with $v\neq 0$.

The results of the numerical analysis is shown in figures 1 - 5,
for some values (including zero) of the curvature $r$. 
We set all bare non-gaussian couplings to zero. 
The bare value of $\lambda_k$ and $\eta_k$
are kept fixed respectively to $u_{40}(t=0) = 0.1$ and $z_{00}(t=0)=1$,
while we have considered 
several bare values of the conformal coupling. We locate the critical 
line for $r=0$ near $u_{20}(t=0)=-0.000325$ and this value 
does not depend on the value of $u_{21}$. 

When the theory is away from the critical line, one sees that 
below a certain value of $t$, the evolution of all 
couplings stops. 
In fact below that scale $f_t$, defined in the appendix,
exponentially vanishes and the tree level scaling 
of the couplings is recovered. In particular 
for the mass 
\be\label{tres}
u_{20}~k^2 = m_k^2 \simeq m^2_R.
\ee
The specific constant value of the renormalized mass $m_R$
depends on how the other parameters  are fixed.
For instance, for the above initial conditions we see that the mass scale
corresponds to $t\sim 6$.

Let us consider the scaling above the mass threshold. 
The running of some non-gaussian couplings is characterized by the  
presence of a ``plateau'' as it is shown in fig.1 for the running of 
$u_{03}=\epsilon_3 k^2$
and we also observe a similar behavior for 
$u_{41}=\zeta_k k^2$. 
This is interesting, we believe, because it 
contrasts with the usual notion  of ``irrelevance'' since those 
interactions behave like marginals for
almost three orders of magnitude below the cutoff. 

A striking result concerns the running of the $u_{22}=\chi_k k^2$ 
interaction term. In fig.2 we see that across and above
the mass scale its value is not negligible, since for instance
$\chi_k k^2 \sim 0.2$ around $t=-4$. Note the strong dependence
of $\chi_k$ from the bare value of $\xi_k$.  

The conformal coupling $\xi_k=u_{21}$ deserves a more detailed discussion. 
Its evolution is shown in fig.3 for the minimally coupled bare theory
and in fig.4 for the conformally coupled bare theory. 
It is worth to remark the small increase of 
$u_{21}$ in fig.3, even for non vanishing non-renormalizable
couplings (see curve 4), within the three orders of magnitude range between 
$t=0$ and the mass threshold, below which the curves become flat.
This means that starting at $t=0$ with a bare $\xi$ far from the 
conformal value 1/6, it is very likely that the coupling does not reach 
that value, being stopped before by the mass threshold.

Actually, looking at the evolution equation for $u_{21}$ in the appendix,
we see that in the infrared region, where the various scales cannot be 
neglected when compared to $k$, it is extremely difficult to establish 
the existence of a non trivial IR fixed point for this coupling.

In fig.4 the behavior of $u_{21}$ in the neighbourhood of 1/6 is explored.
Here we see that there is no deviation from 1/6 only for zero 
non-renormalizable couplings at $t=0$. However, turning on these
couplings, we observe small deviations, of the same order of
magnitude of the changes observed in fig.3.
Again no strong attraction toward $u_{21}=1/6$ is observed.

Finally in fig.5 the influence of the 
curvature on the phase diagram for the $\Phi$ field is investigated.
For a small value of the curvature $r=0.01$, we see that for sufficiently 
high values of the conformal coupling (greater than about 2 in our example)
the running adimensional mass $u_{20}(t)$ grows negative in the IR region
while choosing smaller or negative $u_{21}(t=0)$, 
we see that $u_{20}(t)$ is driven away from the critical line toward 
positive values, at earlier ``times'' $t$. 
In fact, for such curvature, the mass in fig.5 practically represents the 
second
derivative of the potential with respect to the scalar field, since the
additional term $(\xi_k -\eta_k/6)r$ is practically $t$-independent and 
negligible for large values of $t$. 
Thus we are led to a result which is opposite to 
the naive statement, obtained by looking at the not RG-improved propagator,
that the phase transition, when 
$r$ is positive, is approached by decreasing the conformal coupling.

\section{summary and conclusions}
In this paper we have discussed the renormalization of a scalar theory
coupled to the gravitational field by means of the blocking 
procedure. We have obtained non-perturbative flow equations for 
the blocked potential and the wavefunction renormalization 
that rule the flow of all coupling constants generated 
when the cut-off is lowered from the UV region towards
the infrared domain. We have seen that the gravitational field 
influences the scaling laws of the field in the crossover region
between the mass gap and the cut-off. In fact our equations reproduce
the standard perturbative $\beta$-functions only in the UV region
when the contribution of the irrelevant couplings is neglected.  
However we show that any coupling of the kind $R^i \Phi^j$ 
is generated in both potential and wavefunction renormalization 
even if it is not present at the bare level. We have seen that
they can be important in determining the scaling law across the mass gap.
In particular, due to these non renormalizable couplings 
one cannot claim that $\xi=1/6$ is a fixed point for the 
conformal coupling constant whereas, independently on the bare value of 
$\xi_k$, only very small changes of the value of this coupling
are observed  in the flow from the UV to the IR region. 

Another interesting issue addressed with our formalism
concerns the structure of the phase diagram for the quantum field. 
In the model analyzed in the previous section the onset of criticality 
is determined by taking into account the coupling with
the gravitational field. It turns out that the approach to the critical 
line strongly depends on the strenght of the conformal coupling.

It would be important to understand the consequences of our findings
in the contex of the inflationary models. For instance it has been 
recently pointed out that the  ``fine tuning'' problem
in models with the symmetry-breaking scale
near the Planck mass, can be avoided in potentials
dominated by non-quadratic ``irrelevant'' 
interactions terms like $\phi^m$ with $m>2$.
 
In fact in the standard slow-rolling approximation it is 
assumed that the inflaton field evolves as a free field 
leading to a gaussian spectrum for the 
large scale angular anisotropy. Since this happens above the 
mass scale of the inflaton field, one may think that the presence
of non-quadratic interaction terms whose fluctuations are not 
suppressed above the mass scale generates non-gaussian
fluctations in the CMBR. At the present state of the experimental
data there is no secure bound on non-gaussian fluctuations
on large angular scales, therefore this possibility cannot be
ruled out. 
It would be worth to pursue these investigations in a specific model 
in order to answer these questions in detail. 

\acknowledgments
The authors are grateful to Janos Polonyi for his constant advice and many 
important suggestions. A.B has also benefited from discussions with 
M. Reuter and J.Bartlett. A.B.  gratefully acknowledges Fondazione
Angelo Della Riccia and I.N.F.N for financial support. 

\section{appendix}
In this appendix the complete set of equations ruling the couplings of the
action (\ref{3.1}), is displayed in terms of dimensionless quantities introduced
in Eqs. (\ref{2.1},\ref{poze}) for the vacuum characterized by the values 
$\langle \phi(x)\rangle=0$ and  
$R={\cal R}$ or, in terms of dimensionless quantities,
$x=0$ and $y=r$.
Defining $f_t$ as
\be\label{eff} 
f_t^{-1}={u_{22}{r^2}/8}+u_{20}+(u_{21}-{z_{00}/6}){r}
+z_{00}+z_{01}{r}(1-{{r}/ 6})
\ee
we get
\bea\label{nuovo}
&&{dz_{00}\over dt}=2{r}z_{01}-
{1\over 16\pi^2}f_t z_{20}\nonumber\\[2mm]
&&{dz_{20}\over dt}=2z_{20}+
{5\over 16\pi^2}z_{20}f_t^2 (u_{40}+u_{41}{r}+z_{20}(1-{r}/3))
+{1\over 8\pi^2}f_t^3(u_{40}+u_{41}{r}+z_{20}(1-{r}/3))\cdot
\nonumber\\[2mm]
&&(z_{00}+{r}z_{01})
[f_t(z_{00}+{r}z_{01})
(u_{40}+u_{41}{r}+z_{20}(1-{r}/3))
-{r}u_{41}-u_{40}-z_{20}(2-{r}/3)]
\nonumber\\[2mm]
&&{dz_{01}\over dt}=2z_{01}+{1\over 16\pi^2}f_t^2 z_{20}
(u_{21}-{z_{00}\over 6}+{u_{22}{r}\over 4}
+z_{01}(1-{r}/3))\nonumber\\[2mm]
&&{du_{20}\over dt}= -2u_{20}+u_{22}{{r}^2\over 4}
-{f_t\over 16\pi^2}(u_{40}+u_{41}{r}+z_{20}(1-{r}/3))
\nonumber\\[2mm]
&&{du_{21}\over dt}=u_{22}{{r}\over 2}+
{f_t\over 16\pi^2}(z_{20}/3-u_{41})+{f_t^2\over 16\pi^2}
(u_{40}+u_{41}{r}+\nonumber\\[2mm]
&&z_{20}(1-{r}/3))
(u_{21}-{z_{00}\over 6}+{u_{22}{r}\over 4}
+z_{01}(1-{r}/3))\nonumber\\[2mm]
&&{d u_{40}\over dt}=2{r}u_{41}-{f_t\over 16\pi^2}u_{60}
+{3f^2_t\over 16\pi^2}
(u_{40}+u_{41}{r}+z_{20}(1-{r}/3))^2\nonumber\\[2mm]
&&{d u_{60}\over dt}=2u_{60}+{15\over 16\pi^2}f_t^2 u_{60}
(u_{40}+u_{41}{r}+z_{20}(1-{r}/3))-\nonumber\\[2mm]
&&{30\over 16\pi^2}f_t^3
(u_{40}+u_{41}{r}+z_{20}(1-{r}/3))^3\nonumber\\[2mm]
&&{d u_{22}\over dt}=u_{22}/2+
{f_t^2\over 16\pi^2}(u_{40}+u_{41}{r}+z_{20}(1-{r}/3))
(u_{22}/4-z_{10}/3)+\nonumber\\[2mm]
&&{f_t^2\over 8\pi^2}(u_{21}-{z_{00}\over 6}+{u_{22}{r}\over 4}
+z_{01}(1-{r}/3))[(u_{41}-z_{20}/3)-\nonumber\\[2mm]
&&f_t(u_{40}+u_{41}{r}+z_{20}(1-{r}/3))
(u_{21}-{z_{00}\over 6}+{u_{22}{r}\over 4}+z_{01}(1-{r}/3))]\nonumber\\[2mm]
&&{d u_{41} \over dt} = 2 u_{41}+{3f_t^2\over 8\pi^2}
(u_{40}+u_{41}{r}+z_{20}(1-{r}/3))(u_{41}-z_{20}/3)+\nonumber\\[2mm]
&&{f_t^2\over 16\pi^2}(z_{21}-{z_{00}\over 6}+{z_{22}{r}\over 4}+
z_{01}(1-{r}/3))[u_{60}-6f_t(u_{40}+u_{41}{r}+
z_{20}(1-{r}/3))^2]\nonumber\\[2mm]
&&{d u_{01}\over dt}=-2u_{10}-{f_t\over 16\pi^2}(u_{21}-{z_{00}\over 6}+
{u_{22}{r}\over 4}+z_{01}(1-{r}/3))\nonumber\\[2mm]
&&{d u_{02}\over dt}=2u_{03}{r}+{f_t\over 16\pi^2}({z_{10}\over 3}-
{u_{22}\over 4})
+{f_t^2\over 16\pi^2}(u_{21}-{z_{00}\over 6}+{u_{22}{r}\over 4}+
z_{01}(1-{r}/3))^2\nonumber\\[2mm]
&&{d u_{03}\over dt}=4 u_{30}-{4 f_t^3\over 16\pi^2}
(u_{21}-{z_{00}\over 6}+{u_{22}{r}\over 4}+
z_{01}(1-{r}/3))^3-\nonumber\\[2mm]
&&{6f_t^2\over 16\pi^2}({z_{10}\over 3}-{u_{22}\over 4})
(u_{2,1}-{z_{0,0}\over 6}+{u_{22}{r}\over 4}+
z_{0,1}(1-{r}/3))\nonumber\\[2mm]
&&{d u_{00}\over dt}=-4u_{00}-2u_{01}{r}+{u_{03}\over 3}{r}^3
+{1\over 16\pi^2}\ln {f_t\over f_t(x=0, y=r)}.
\eea

\newpage
\begin{center}
FIGURE CAPTIONS
\end{center}
\vskip 30 pt

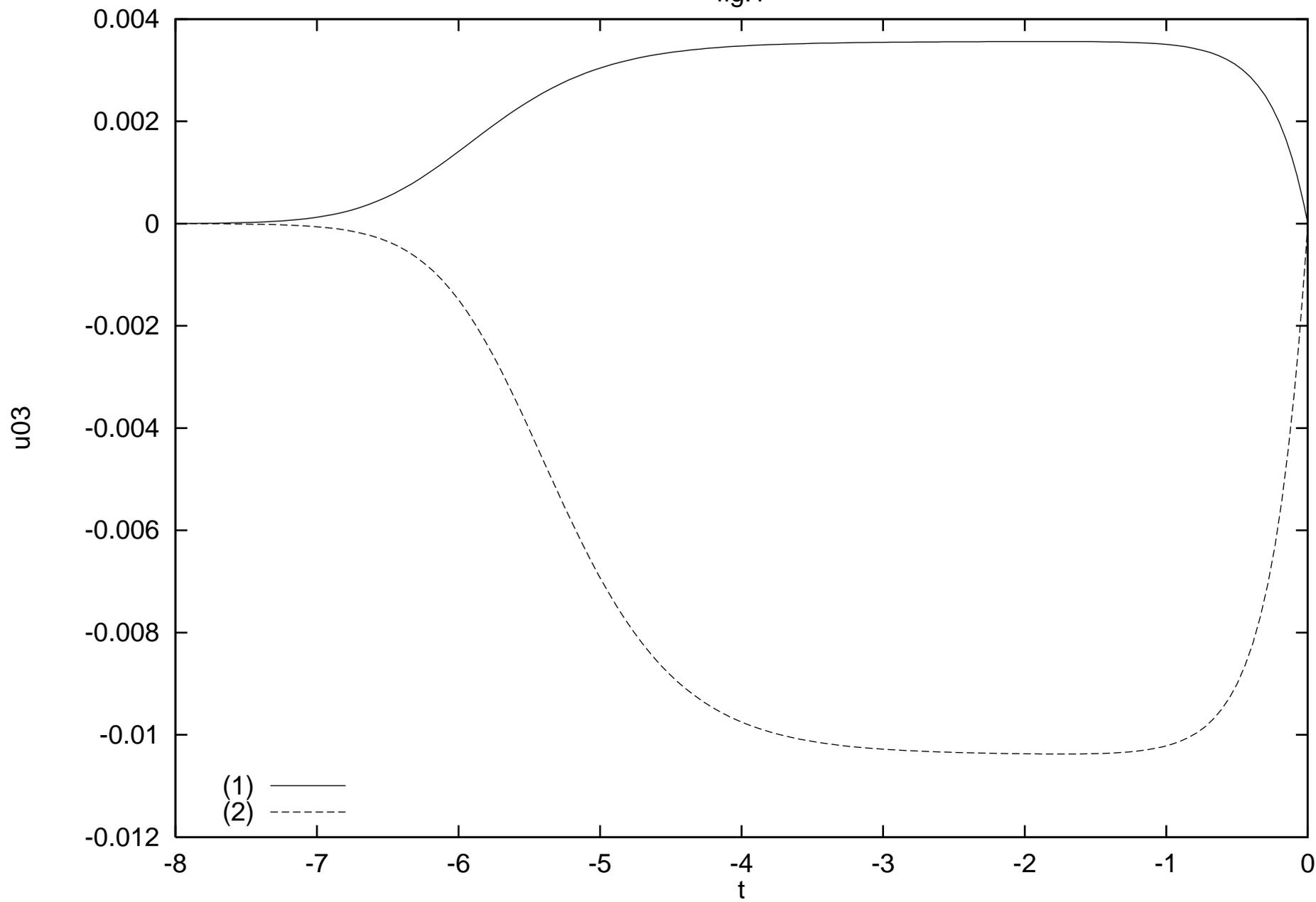
Figure 1. $u_{03}(t)$ with 
$u_{21}(t=0)=1$, (1), and
$u_{21}(t=0)=-1$, (2). For both curves the non renormalizable couplings 
are fixed to zero at $t=0$ and $r=0.01$.

\vskip 20pt

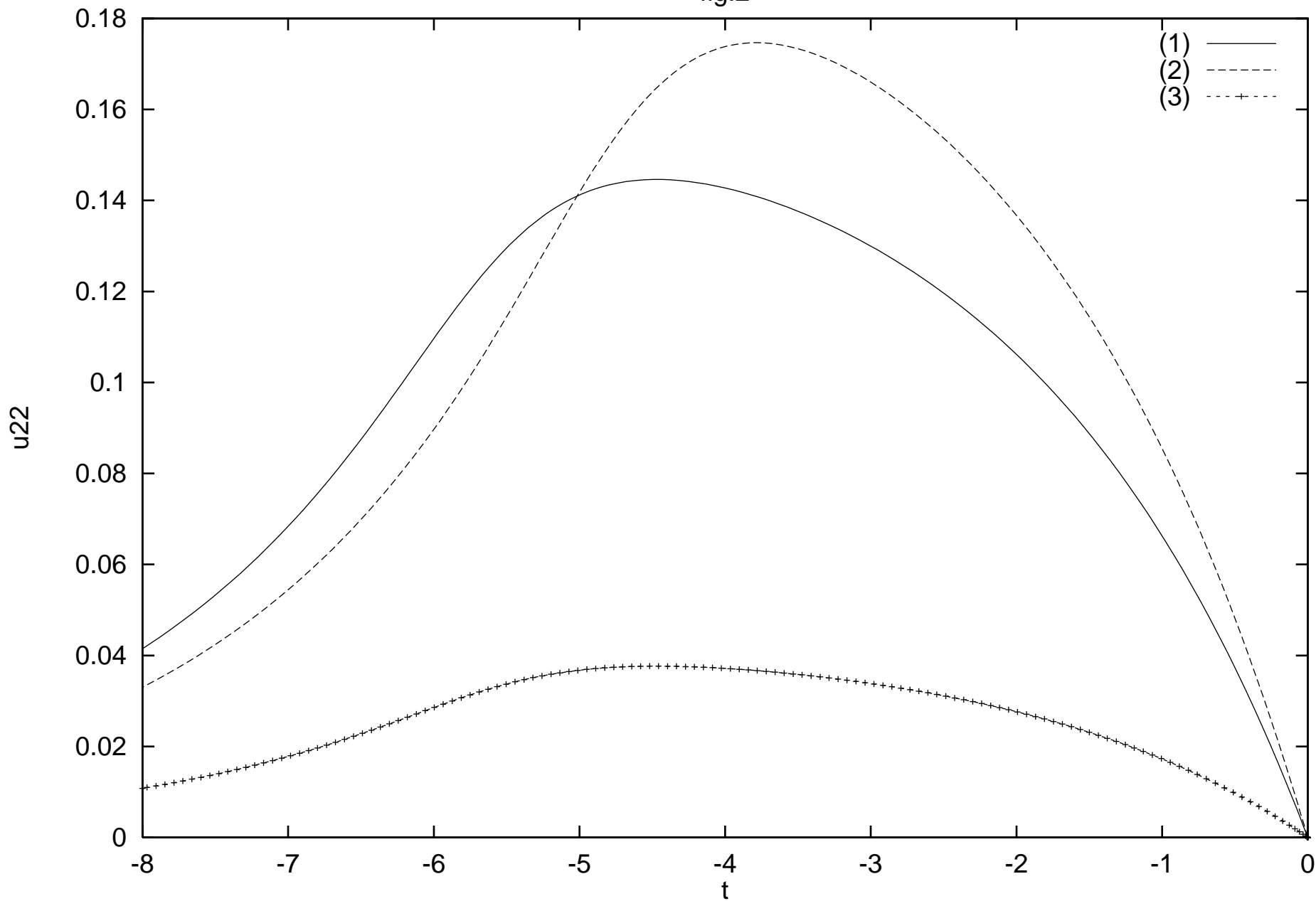
Figure 2. $u_{22}(t)$ with 
$u_{21}(t=0)=-8$ and $r=0$ (1),
$u_{21}(t=0)=-8$ and $r=0.01$ (2),
$u_{21}(t=0)=-4$ and $r=0.1$ (3).
The non renormalizable couplings are fixed to zero
at $t=0$ for all curves.

\vskip 20pt

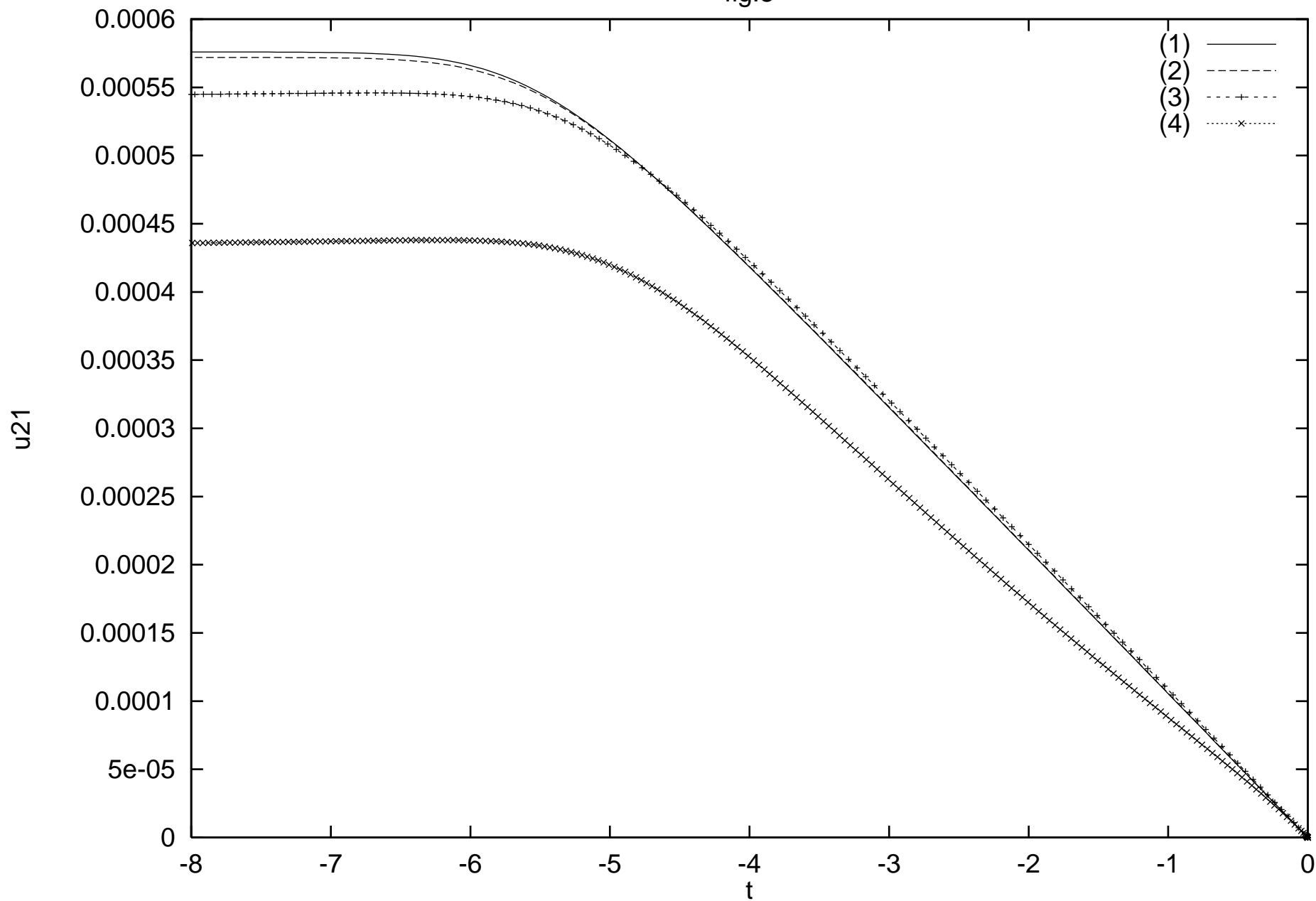
Figure 3. Flow of $u_{21}(t)$ starting at  $u_{21}(t=0)=0$
for zero non-renormalizable couplings at $t=0$ and
$r=0$ (1), $r=0.01$ (2), $r=0.1$ (3);
with the various non-renormalizable couplings 
set to $0.01$ at $t=0$ and $r=0.01$ (4).

\vskip 20pt

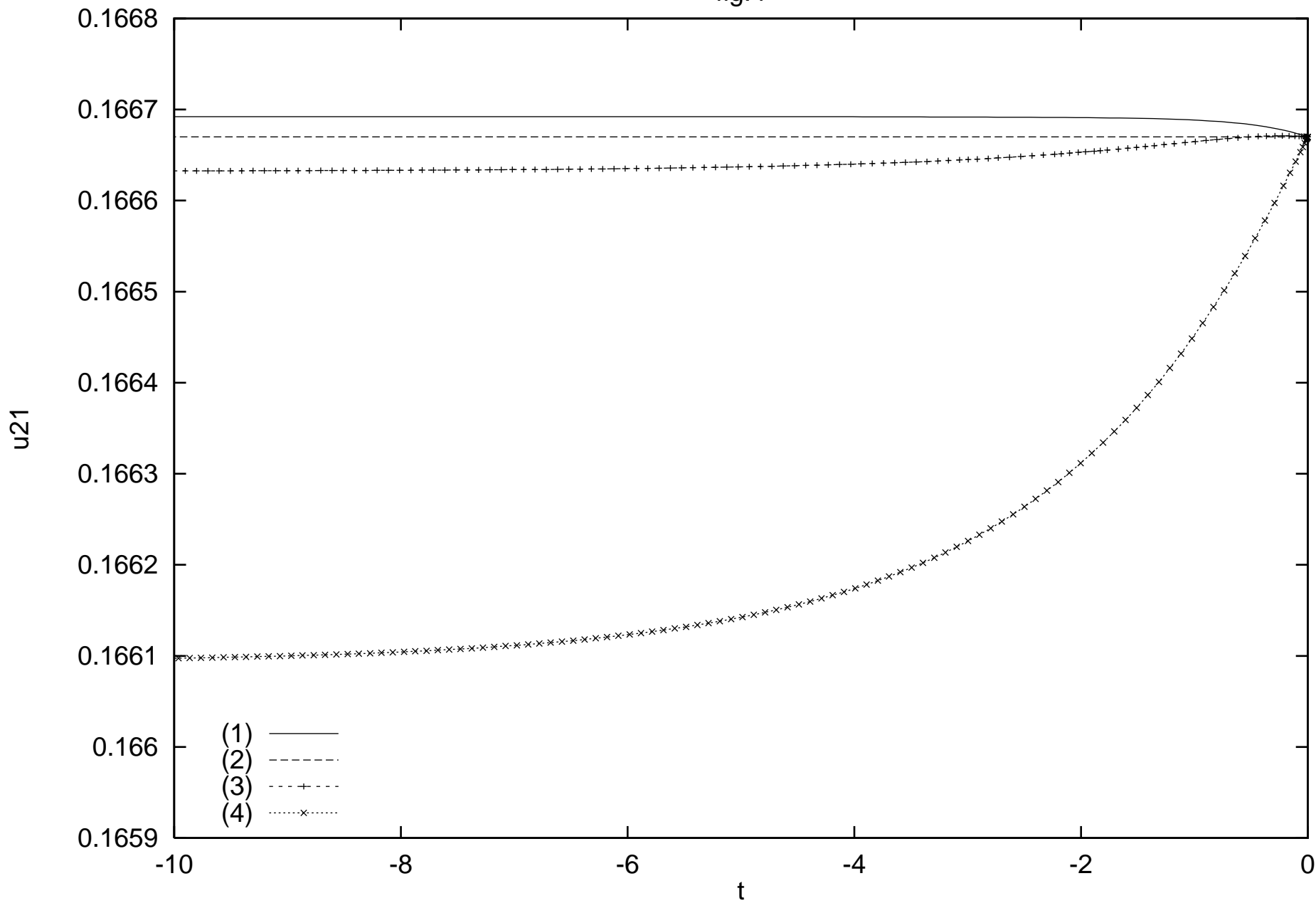
Figure 4. Flow of $u_{21}(t)$ starting at  $u_{21}(t=0)=1/6$
with the various non-renormalizable couplings 
set to $0.01$ at $t=0$ and and $r=0$ (1),
$r=0.01$ (3), $r=0.1$ (4). The flat curve (2)
corresponds to zero non-renormalizable couplings at $t=0$ and it is not 
sensitive to $r$.

\vskip 20pt

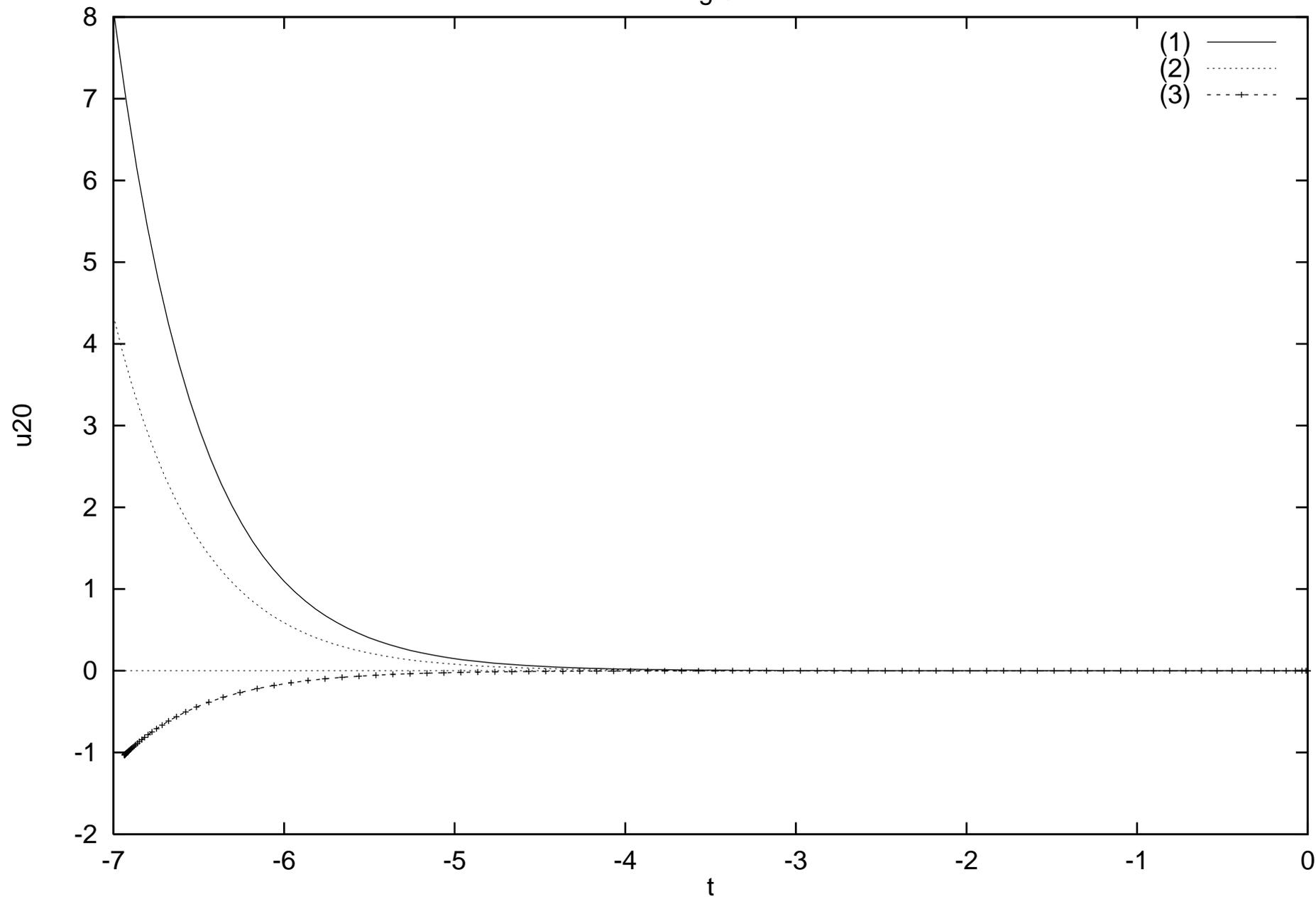
Figure 5. $u_{20}(t)$ plotted 
for three different values of the conformal coupling:
$u_{21}(t=0)=0$ (1), 
$u_{21}(t=0)=1$ (2), 
$u_{21}(t=0)=2.5$ (3);  $r$ is kept fixed to $r=0.01$ and
the the bare non-renormalizable couplings are zero.

\end{document}